\documentclass[a4paper,10pt]{cpc-hepnp}
\usepackage{multicol}
\usepackage{graphicx}
\usepackage{amssymb,bm,mathrsfs,bbm,amscd}
\usepackage{booktabs}
\usepackage[tbtags]{amsmath}
\usepackage{lastpage}

\begin{document}

\fancyhead[co]{\footnotesize Wang Mei-Juan~ et al:  Eccentricity and
elliptic flow at fixed centrality in Au+Au collisions at
$\sqrt{s_{\rm NN}}$=200GeV  }

\title { Eccentricity and elliptic flow at fixed centrality in Au+Au collisions at
$\sqrt{s_{\rm NN}}$=200GeV in AMPT model
\thanks{\footnotesize {Supported by $'$the Fundametal Research Funds for Central Universities$'$ (GUGL 100237) and National Natural Science
Foundation of China (10835005).}}}

\author{%
WANG Mei-Juan$^{1; 1)}$\email{wangmj@iopp.ccnu.edu.cn}%
\quad CHEN Gang$^{1}$%
\quad WU Yuan-Fang$^{2}$ }

\maketitle

\address{
$^1$ Physics Department, China University of Geoscience, Wuhan 430074,
China\\
$^2$ Institute of Particle Physics, Huazhong Normal University, Wuhan
430079, China}

\begin{abstract}
In this paper, elliptic flow is studied at fixed centrality in Au+Au
collision at $\sqrt{s_{\rm NN}}$=200GeV in the AMPT model. It is
observed that with the participant increasing, elliptic flow has an
increase or a decrease at different fixed impact parameter, but it
does not have a trivial fluctuation. It is analyzed that the initial
space anisotropy dominates the participant dependence of elliptic
flow in near-central collisions(b=5fm) and mid-central
collisions(b=8fm), while the interaction between particles can
mainly answer for the behavior of elliptic flow with participant in
peripheral collisions(b=12fm). To distinguish the pure geometrical
effect, elliptic flow scaled by initial eccentricity is studied. It
is found that the ratio $v_{2}/\epsilon$ increases with participant
and reaches a saturation when the participant is large enough,
indicating that the collision system may reach the local
equilibrium.
\end{abstract}



\begin{multicols}{2}

\section {Introduction}
The discovery of a large azimuthal anisotropic flow of hadrons at
RHIC provides a conclusive evidence for the created dense partonic
matter in ultrarelativistic nucleus-nucleus
collisions~\cite{qgp0}~\cite{qgp1}~\cite{qgp2}. The strong
interaction medium in the collision zone can be expected to achieve
a local equilibrium and exhibit an approximately hydrodynamics flow
~\cite{emit}~\cite{Heinz}~\cite{Shuryak}. Moreover, the study of
anisotropic flow has a potential to offer insights into the equation
of state of the produced matter~\cite{Eos1}~\cite{Eos2}.

The momentum anisotropy of final particles is generated due to the
transverse density gradient based on an initial geometry of an
$"$almond-shaped$ "$ overlap region produced in non-central
collisions. The pressure gradient converts the initial coordinate
space asymmetry into the momentum anisotropy of final particles,
such as elliptic flow. The magnitude of the elliptic flow depends on
both initial spatial asymmetry in non-central collisions and the
subsequent interaction between the particles. Therefore, the study
of elliptic flow is very crucial to understand the properties of the
dense matter formed during the initial stage of heavy ion
collisions~\cite{initial} and parton dynamics~\cite{dynamics} at the
relativistic heavy ion energies.

In this paper, we focus on the study of elliptic flow at fixed
centrality where the impact parameter b is a constant. The elliptic
flow at fixed centrality in Au+Au collisions at $\sqrt{s_{\rm
NN}}$=200GeV is presented in detail, such as the variation with the
number of participating nucleons and the behavior after scaled by
the initial eccentricity. Here, the model we used is the AMPT with
string melting~\cite{ampt_sum}.

The layout of this paper is in the following. Section 2 briefly
introduces the AMPT model, a transport model based on parton level.
In Section 3, we mainly study the elliptic flow and initial
eccentricity at fixed centrality in Au+Au collisions, the dependence
of participating nucleons, the ratio $v_{2}/\epsilon$, and so on.
Finally, a conclusion is given in Section 4.

\section{ A brief introduction to AMPT}

The AMPT model~\cite{ampt_sum} is based on parton level transport
dynamics. There are two versions of AMPT model, one is the default
AMPT, and the other is AMPT with string melting. It turns out that
the default AMPT (v1.11) is able to give a reasonable description on
hadron rapidity distributions and transverse momentum spectra
observed in heavy ion collisions at both SPS and RHIC. However, it
fails to reproduce the experimental data about elliptic flow and
two-pion correlation function. On the other hand, the AMPT model
with string melting (v2.11) can well describe the elliptic flow and
two-pion correlation function~\cite{ampt_flow}~\cite{ampt_hbt} but
agrees bad with the hadron rapidity and transverse momentum spectra.
According to our demand, the AMPT with string melting is chosen.

The AMPT model with string melting contains four main components:
the initial conditions, partonic interactions, conversion from the
partonic to the hadronic matter and hadronic interactions. The
initial conditions are obtained from the HIJING model~\cite{hijing}.
The time evolution of partons is then modeled by the ZPC parton
cascade model~\cite{zpc}. After partons stop interacting, a simple
quark coalescence model is used to combine the two nearest quarks
into a meson and three nearest quarks (antiquarks) into a baryon
(antibaryon). After hadronization, scatterings among the resulting
hadrons are described by a relativistic transport (ART)
model~\cite{art} which includes baryon-baryon, baryon-meson and
meson-meson elastic and inelastic scatterings.

In the following we will utilize the AMPT with string melting to
generate Au+Au collision events at $\sqrt{s_{\rm NN}}=200$ GeV. The
parton cross section is taken to be 10 mb.

\section{ The elliptic flow, initial eccentricity and their ratio $v_{2}/\epsilon$ }

In Section 3.1, we will firstly study the elliptic flow in Au+Au
collisions from minibias events, and it is observed that the
strongest collective behavior appears in the mid-central collisions.
In Section 3.2, we present the participant dependence of elliptic
flow and initial eccentricity at fixed centrality. Finally, the
participant dependence of the ratio $v_{2}/\epsilon$ is shown in
Section 3.3.

\subsection{ The elliptic flow from minibias events}
When two nuclei collide at nonzero impact parameter, their overlap
in the transverse plane has a short axis, parallel to the impact
parameter, and a long axis perpendicular to it. This initial space
anisotropy is converted by the pressure gradient into a momentum
asymmetry, so more particles are emitted along the short
axis~\cite{emit}. This magnitude of this effect is characterized by
the elliptic flow, defined as

\begin{eqnarray}
\label{defv2} v_2 =\langle\cos 2(\varphi-\Phi_R)\rangle,
\end{eqnarray}

where $\varphi$ is the azimuthal angle of an outgoing particle,
$\Phi_R$ is the azimuthal angle of the impact parameter,  and
angular brackets denote an average over many particles and many
events.

In this section, we study the elliptic flow signal as a function of
the impact parameter b(Fig1.(a)), the initial
participants(Fig1.(b))and the final charged particles(Fig1.(c)) in
Au + Au collisions at $\sqrt{s_{\rm NN}}$=200GeV from the AMPT model
with string melting. Here, all the charged particles with rapidity
in the region y$\in$(-5,+5) are included.

From the figure we can see that the elliptic flow first increases,
reaches its maximum value and then decreases, indicating the
strongest collective behavior is produced in mid-central Au+Au
collisions. The elliptic flow is built and formed from the
anisotropic geometrical effect and the hadron and parton interaction
between particles during the system expansion. In peripheral
collisions, we can obtain the strongest initial anisotropy, while
the interaction between produced particles is weak when the number
of participants is so small. In the near-central collisions, most
nucleons can take part in the collision, while the initial
anisotropy is not large enough. It is reasonable that that the
strongest elliptic flow appears in the mid-central collision where
the interaction between particles is strong enough to convert the
initial space anisotropy into final momentum anisotropy completely.

\begin{center}
\includegraphics[width=3in]{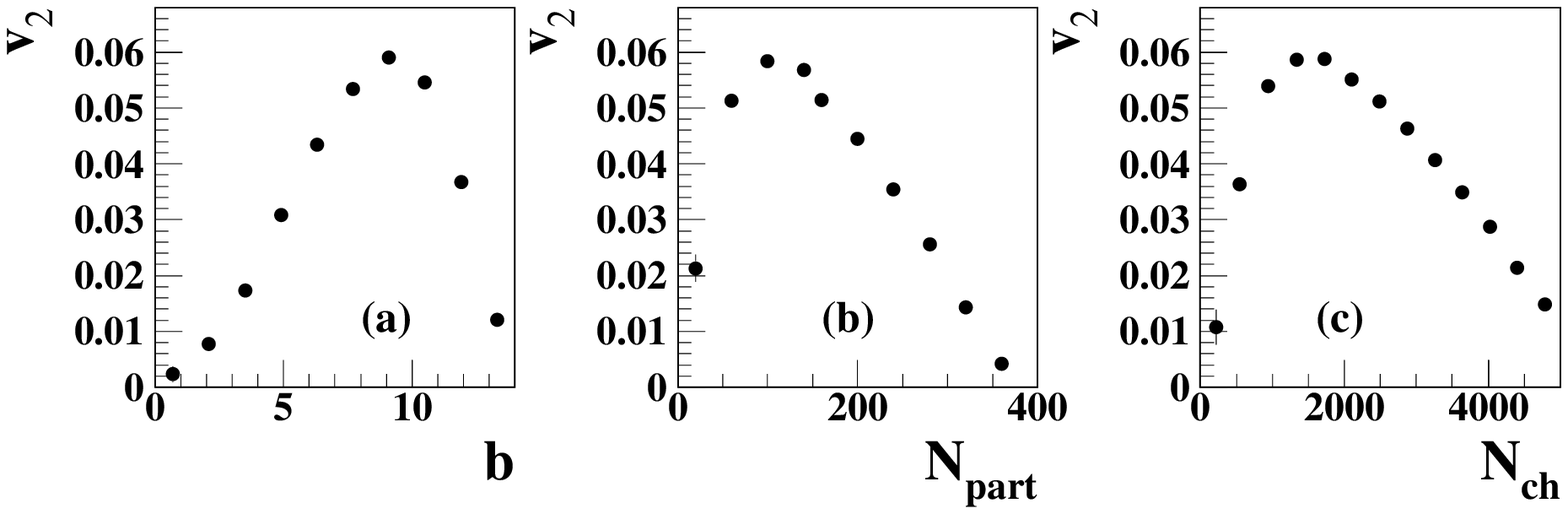}
\figcaption{\small \label{Fig.1} The impact parameter (a), the
participant (b) and the final charged particles (c) dependence of
elliptic flow in Au + Au collisions at 200 GeV with AMPT model.}
\end{center}

\subsection{ The participant dependence of the elliptic flow and initial eccentricity at fixed centrality}
In this section, we focus on the study of the elliptic flow at fixed
centrality where the impact parameter b is a constant. Here, we
choose the fixed impact parameter b =5fm, 8fm and 12fm, which
responds to the case in the near-central, the mid-central and the
peripheral collisions respectively. In general, the geometric
overlap region of Au+Au collisions is fixed, elliptic flow is
expected to remain roughly unchanged or at most have a trivial
fluctuation.

The participant dependence of elliptic flow in Au+Au collisions at
fixed centrality at$\sqrt{s_{\rm NN}}$=200GeV is shown in the upper
panel of Fig. 2. From the figure, we can see that even if the impact
parameter b is fixed, the fluctuation of participating nucleons is
so large that it can't be ignored. With the participant increasing,
the elliptic flow has a slow increase in the peripheral collisions
for b =12fm (left), while decreases monotonously in the  mid-central
and near-central collisions, e.g., b=8fm (middle) and b=5fm (right).
As a result, it is argued that the participant dependence of
elliptic flow at fixed centrality has some physics origin, but not
the only statistical fluctuation effect.

\begin{center}
\includegraphics[width=3in]{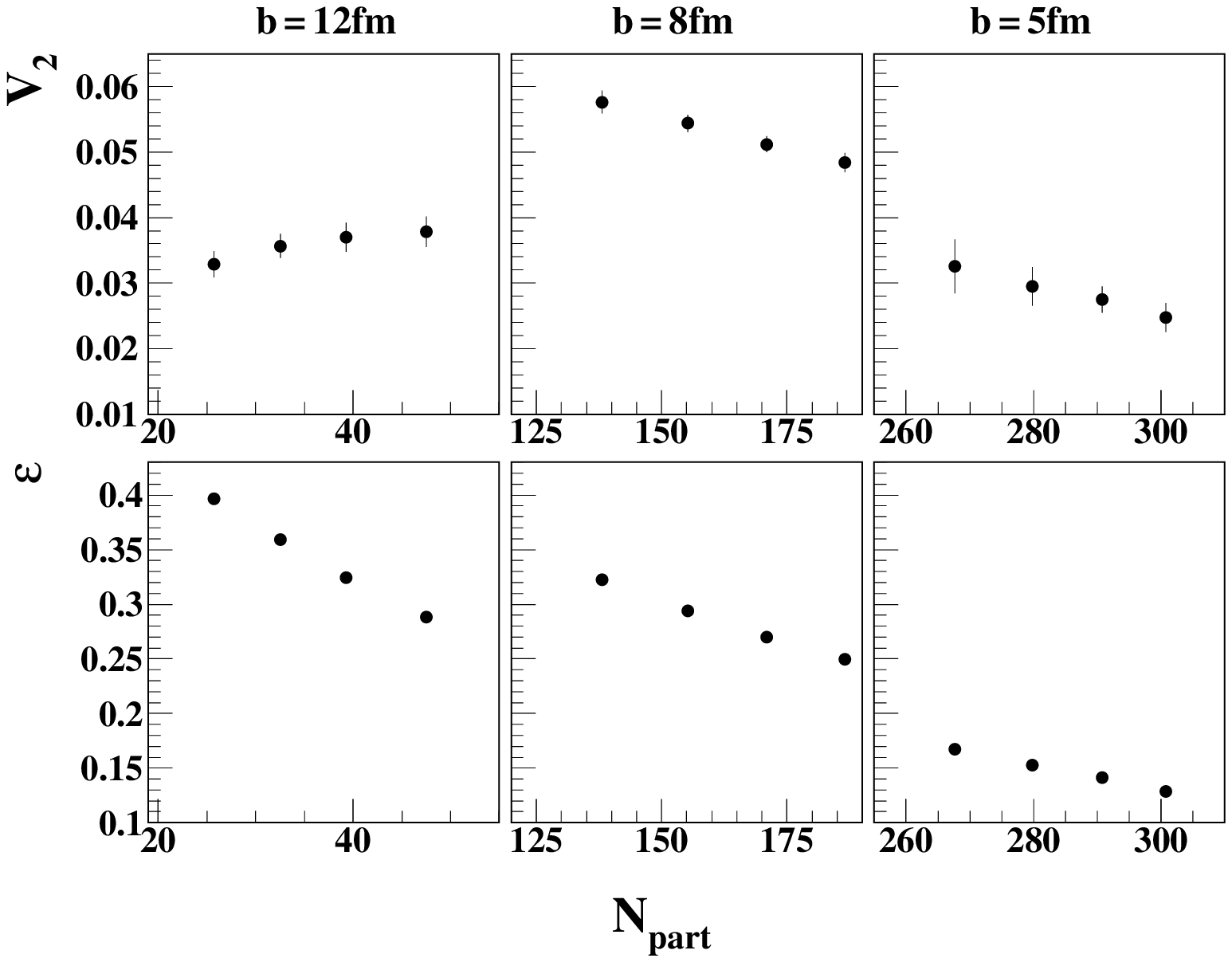}
\figcaption{\small \label{Fig. 2} The participant dependence of
elliptic flow(the upper panel) and eccentricity (the lower panel)
for 200 GeV Au+Au collisions in AMPT with string melting at fixed
impact parameter b=12fm (left), b=8fm (middle) and b=5fm (right).}
\end{center}

It is widely believed that the anisotropic collective flow is mainly
driven by the initial eccentricity of the matter created in nuclear
overlap zone. In this point, we will study simply the participant
dependence of the corresponding initial eccentricity $\epsilon$ at
fixed centrality. The Monte Carlo Glauber model~\cite{Glauber} is
used to estimate the initial eccentricity from the distribution of
participant nucleons in the transverse plane.  The participant
eccentricity is defined by~\cite{definition-e}
 \begin{eqnarray}
{\rm \varepsilon_{\rm part} =
\frac{\sqrt{(\sigma_{y}^2-\sigma_{x}^2)^2+4\sigma_{xy}^2}}{\sigma_{x}^2+\sigma_{y}^2},
 \label{eqeccpart}}
 \end{eqnarray}
where $\sigma_{x}^2=\langle x^2\rangle - \langle x\rangle^2$ and
$\sigma_{y}^2=\langle y^2\rangle - \langle y\rangle^2$ are the
variances of the nucleon distribution in the x- and the y-direction,
and $\sigma_{xy}=\langle xy\rangle - \langle x\rangle\langle
y\rangle$ is the covariance of the position of participant nucleons
.

The participant dependence of initial eccentricity at fixed
centrality in Au+Au collisions at $\sqrt{s_{\rm NN}}$=200 GeV is
presented in the lower panel of Fig. 2. From the figure, we can see
the initial eccentricity $\epsilon$ for the cases of b=12fm (left),
8fm (middle) and 5fm (right), all have a decrease with the
participant increasing.

Comparing the results of the upper and down panel in Fig. 2,   it is
obvious that the elliptic flow and the initial eccentricity have a
consistent decrease with the increasing participant both for b=8fm
(middle) and b=5fm (right), while they have the contrast case for
b=12fm (left). As we know, the elliptic flow is formed and built
from the initial anisotropy of the overlap region and the subsequent
interaction between the produced particles. On this basis, we argue
that the initial space anisotropy dominates the participant
dependence of elliptic flow in near-central (b=5fm) and mid-central
collisions (b=8fm). The subsequent interaction between the produced
particles, which becomes stronger with the participant increasing,
can mainly answer for the participant dependence of elliptic flow in
peripheral collisions(b=12fm), such as a slow increase.

\subsection{ The participant dependence of $v_{2}/\epsilon$ at fixed centrality}

\begin{center}
\includegraphics[width=3in]{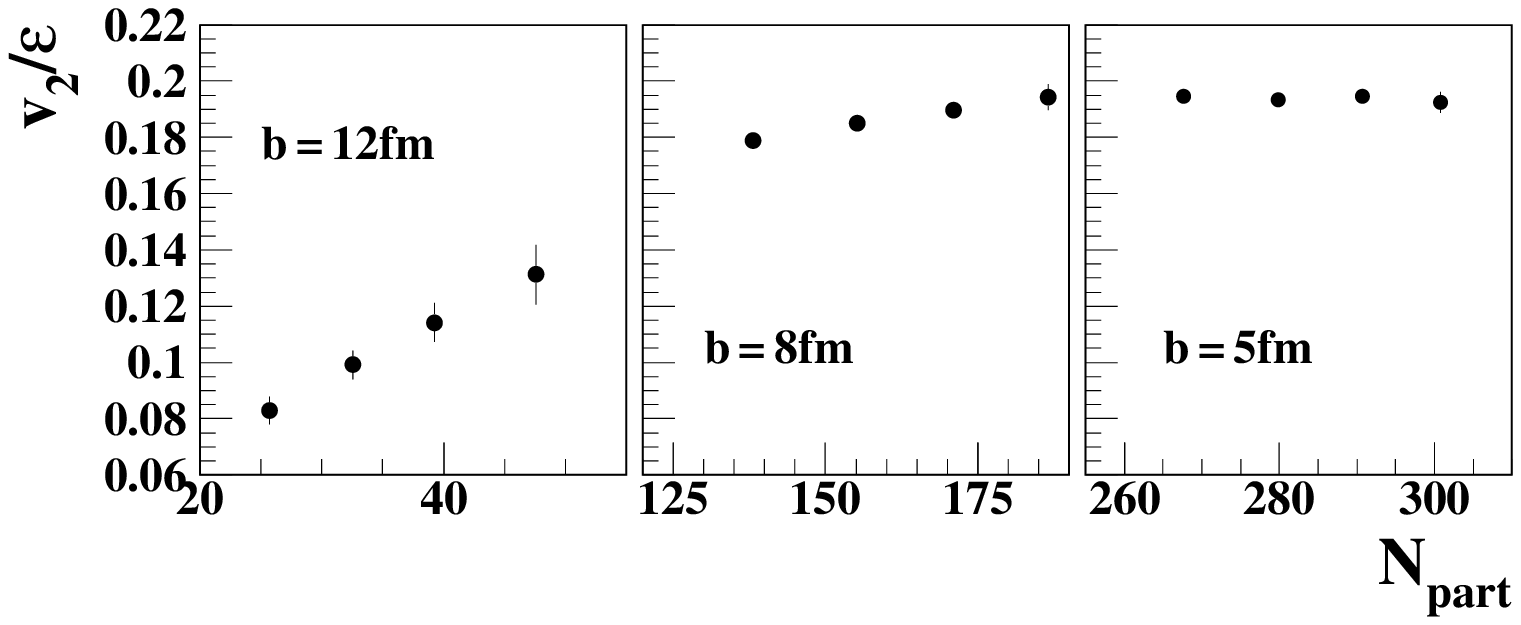}
\figcaption{\small \label{Fig. 3} The participant dependence of the
ratio $v_{2}/\epsilon$ in Au + Au collisions at 200GeV at fixed
impact parameter b=12fm (left), b=8fm (middle) and b=5fm (right) in
the AMPT model with string melting.}
\end{center}

In order to distinguish the collision dynamics from the purely
geometrical effects, it has been suggested that the measured $v_{2}$
should be scaled by the initial eccentricity of the nuclear overlap.
If the produced matter equilibrates, it behaves as an ideal fluid.
Hydrodynamics predicts that $v_{2}$ scaled by eccentricity
$\epsilon$ has a saturation when the collision system achieves the
local equilibrium~\cite{predicition1}~\cite{predicition2}. However,
if equilibration is not incomplete, then the eccentricity scaling is
indeed broken, e.g., in the peripheral Au+Au
collisions~\cite{calculation}~\cite{recent-star}.

The participant dependence of $v_{2}/\epsilon$ at fixed centrality
in Au+Au collisions at $\sqrt{s_{\rm NN}}$=200 GeV is shown in Fig.
3. From the figure, we can see that the ratio $v_{2}/\epsilon$ keeps
increasing both for b=12fm (left) and b=8fm (middle), while it
becomes saturated in the near-central collisions for b=5fm (right).
It is reasonable that more participants lead to stronger interaction
between particles, hence a larger ratio $v_{2}/\epsilon$ can be
obtained in more central collisions. As expected in an equilibrium
scenario, the ratio $v_{2}/\epsilon$ in the near-central collisions
shows little sensitivity to the participants. This indicates that
the system created in 200GeV Au+Au near-central collision by the
AMPT model including both parton cascade and hadron scattering, may
reach local thermalization when the interaction between particles is
strong enough.

\section{Conclusion}

To summarize, we have studied the elliptic flow and initial
eccentricity and their ratio at fixed centrality in Au+Au collisions
at $\sqrt{s_{\rm NN}}=200$ GeV . It is observed that they have a
monotonous behavior(an increase or a decrease) with the participant
increasing for different fixed impact parameter b, but they do not
have a trivial fluctuation. It is argued that the initial
eccentricity dominates the participant dependence of elliptic flow
in the mid-central collisions (b=8fm) and near-central collisions
(b=5fm), while the interaction between particles dominates the
behavior of elliptic flow with the participants in peripheral
collisions (b=12fm). Moreover, it is found that the elliptic flow
scaled with initial eccentricity keeps increasing with the
participants. When the number of participants is large enough, the
ratio $v_{2}/\epsilon$ has a saturation, indicating that the
collision system may reach local equilibrium.
\end{multicols}

\vspace{-1mm}
\centerline{\rule{80mm}{0.1pt}}
\vspace{2mm}

\begin{multicols}{2}

\end{multicols}

\clearpage


\begin{thebibliography}{90}

\vspace{3mm}

\bibitem{qgp0} Marcus Bleicher and Horst St\"{o}cker, Phys. Lett.
B, 2002, {\bf 526}, 309-314.

\bibitem{qgp1} K. Adcox et al.(PHENIX Collaboration),  Nucl.
Phys. A, 2005, {\bf 757}, 184-283.

\bibitem{qgp2} John Adams et al.(STAR Collaboration), Nucl. Phys.
A, 2005, {\bf 757}, 102-183.

\bibitem{emit} J. Y. Ollitrault,  Phys. Rev . D, 1992, {\bf 46},
229.

\bibitem{Heinz} U. Heinz and P. Kolb,  Nucl. Phys. A, 2002, {\bf 702}, 269.

\bibitem{Shuryak} E. Shuryak,  Prog. Part. Nucl. Phys., 2009, {\bf 62}, 48.

\bibitem{Eos1}J. Brachmann, S. Soff, A. Dumitru, H. St\"{o}cker, J.
A. Maruhn, W. Greiner, D. H. Rischke, Phys. Rev. C, 2000,  {\bf 61},
024909.

\bibitem{Eos2} J. H. Chen et al,  Phys. Rev. C, 2006,  {\bf 74}, 064902;
 T. Z. Yan et al,  Phys. Lett. B, 2006,  {\bf 638}, 50.


\bibitem{initial} P. F Kolb, P. Huovinen, U. Heinz and H. Heiselberg,
 Phys. Lett. B, 2001,  {\bf 500}, 232.

\bibitem{dynamics} B. Zhang, M. Gyulassy and Che-Ming Ko,
Phys. Lett. B, 1999, {\bf 455}, 45.

\bibitem{ampt_sum} Zi-Wei Lin, Che Ming Ko, Bao-An Li and Bin Zhang and
Subrata Pal, Phys. Rev. C, 2005, {\bf 72}, 064901.

\bibitem{ampt_flow} Zi-Wei Lin and C. M. Ko, Phys. Rev. C, 2002,  {\bf 65},  034904.

\bibitem{ampt_hbt} Zi-Wei Lin, C.M. Ko and Subrata Pal,
Phys. Rev. Lett, 2002, {\bf 89}, 152301.

\bibitem{hijing} X. N. Wang, Phys. Rev. D, 1991, {\bf 43}, 104; M. Gyulassy and X. N. Wang, Comput. Phys. Commun., 1994,  {\bf 83}, 307.

\bibitem{zpc} B. Zhang, Comput. Phys. Commun., 1998,  {\bf 109}, 193.

\bibitem{art} B. A. Li and C. M. Ko, Phys. Rev. C, 1995,  {\bf 52}, 2037.

\bibitem{Glauber}Michael L. Miller, Klaus Reygers, Stephen J. Sanders, Peter
Steinberg, Ann. Rev. Nucl. Part. Sci. , 2007, {\bf 57}, 205-243.

\bibitem{definition-e} B. Alver et al. (PHOBOS), Phys. Rev. Lett, 2007, {\bf 98}, 242302.

\bibitem{predicition1}J. Y. Ollireault, Phys. Rev. D, 1992, {\bf 46}, 229.

\bibitem{predicition2}H. Sorge, Phys. Rev. Lett, 1999, {\bf 82}, 2048.

\bibitem{calculation} Hans-Joachim Drescher, Adrian Dumitru, Clement
Gombeaud, and Jeans-Yves Ollitrault, Phys. Rev. C, 2005,  {\bf 76},
024905.

\bibitem{recent-star}B. l. Abelev, et al.(STAR), Phys. Rev. C, 2008,  {\bf 77}, 054901.

\end{thebibliography}
\end{document}